\documentclass{iaus}
\usepackage{graphicx}
\usepackage{subfigure}
\usepackage{natbib}

\newcommand{\Msun}{\mbox{$\mathrm{M}_{\odot}$}}

\title[Rotational mixing in close binaries]{Rotational mixing
in close binaries}

\author[S.E. de Mink, M. Cantiello \& N. Langer]
{S.E. de Mink$^1$,
 M. Cantiello$^1$,  N. Langer$^1$, \break S.-Ch. Yoon$^2$, I. Brott$^1$, E. Glebbeek$^1$, M. Verkoulen$^1$,  \and O.R. Pols$^1$}

\affiliation{$^1$Astronomical Institute Utrecht,
              Princetonplein 5, 3584 CC Utrecht, The Netherlands\break           
	      $^2$Dep. of Astronomy \& Astrophysics, Univ. of California, Santa Cruz, CA95064, USA\break
   email: S.E.deMink@uu.nl, M.Cantiello@uu.nl, N.Langer@uu.nl
}

\pubyear{2008}
\volume{252}  
\pagerange{???-???}
\date{?? and in revised form ??}
\jname{Proceedings The Art of Modelling Stars in the 21st century}

\editors{Licai Deng  K.L. Chan \& C. Chiosi, eds.}

\begin{document}

\maketitle

\begin{abstract}

Rotational mixing is a very important but uncertain process in the
evolution of massive stars.  We propose to use close binaries to test
its efficiency.
Based on rotating single stellar models we predict nitrogen surface
enhancements for tidally locked binaries. Furthermore we demonstrate
the possibility of a new evolutionary scenario for very massive
($M>40\Msun$) close ($P<3$~days) binaries: Case M, in which mixing is
so efficient that the stars evolve quasi-chemically homogeneously,
stay compact and avoid any Roche-lobe overflow, leading to very close
(double) WR binaries.


\end{abstract}
\firstsection 

\begin{figure}[b]
\centering
  \includegraphics[width=0.7\textwidth]{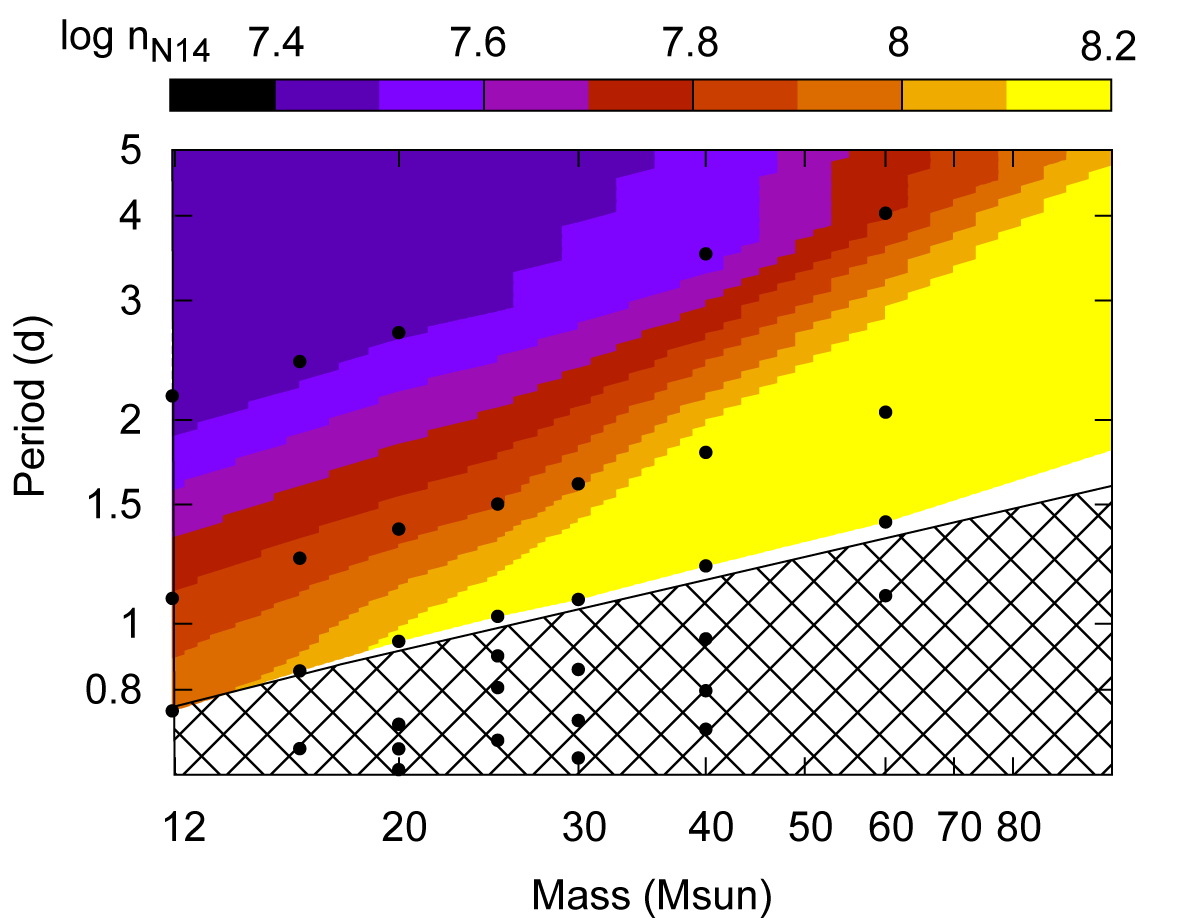}
  \caption{ The nitrogen surface abundance is plotted in color
  shading for rotating single stellar models (black dots) at a
  metallicity of Z$=$0.004 with different masses and initial spin
  periods at the end of their main sequence evolution. In a tidally
  locked binary the spin period corresponds to the orbital period.
  The hashed region is excluded for binaries as the two stars are in
  contact at zero age.  }\label{fig:N}
\end{figure}

\section{Introduction}

Rotation plays an important role in the evolution of massive stars: it
causes deformation of the star due to the centrifugal force, it
interplays with  stellar mass loss, and it can induce instabilities leading
to internal mixing of the star.
%
%
Mixing induced by rotation has been successful in explaining the ratio
of red to blue super giants and it has been invoked to explain the
helium and nitrogen enhancements observed in OB stars \citep[][and
references therein]{Heger+Langer00, Maeder+Meynet00}.
%

Although extensive literature exists on the subject, the
 efficiency of rotationally induced mixing is still very
 uncertain. The VLT-FLAMES survey of massive stars
 \citep[][]{Evans+05}, which resulted in rotational velocities and
 surface abundances of about one thousand O and early B stars provided a
 major step forward. However it raised more questions than it answered
 regarding rotational mixing \citep{Hunter+07}.
This motivated us to formulate potential observational tests to
constrain the corresponding uncertain physical parameters. For this
purpose we focus on tidally locked binaries.

The advantage of using binaries is that the major stellar parameters,
such as the masses, radii and effective temperatures, can be
accurately determined \citep[e.g.][]{hilditch+05} and also, if high
resolution spectra are available, the surface abundances.  This
enables us to test our stellar models directly against well understood systems
\citep[see also][]{demink+07}.
A second advantage of using close binaries is that the tidal forces
synchronize the spin period of the stars with the orbital period, such
that $P_{\rm spin} = P_{\rm orbit}$. This enables us to determine the
rotation rate of the stars much more accurately than in the case of
single stars, where it can only be estimated from $v \sin i$, derived
from spectral fitting. The inclination $i$ of th rotation axis is
generally not known for a particular single star.

In this contribution we use of rotating single stellar models (as
published by \citet{Yoon+06}, to which we refer for details) to
demonstrate two predictions for detached tidally locked binaries. In
Section~\ref{sec:surf} we discuss the surface enhancement which can be
expected in close binaries. In Section~\ref{sec:caseM} we discuss a
new evolutionary binary scenario for the most massive close binaries,
case M, in which rotational mixing is so efficient that the stars stay
compact and avoid any Roche lobe overflow.

\begin{figure}
\centering
  \includegraphics[width=0.7\textwidth]{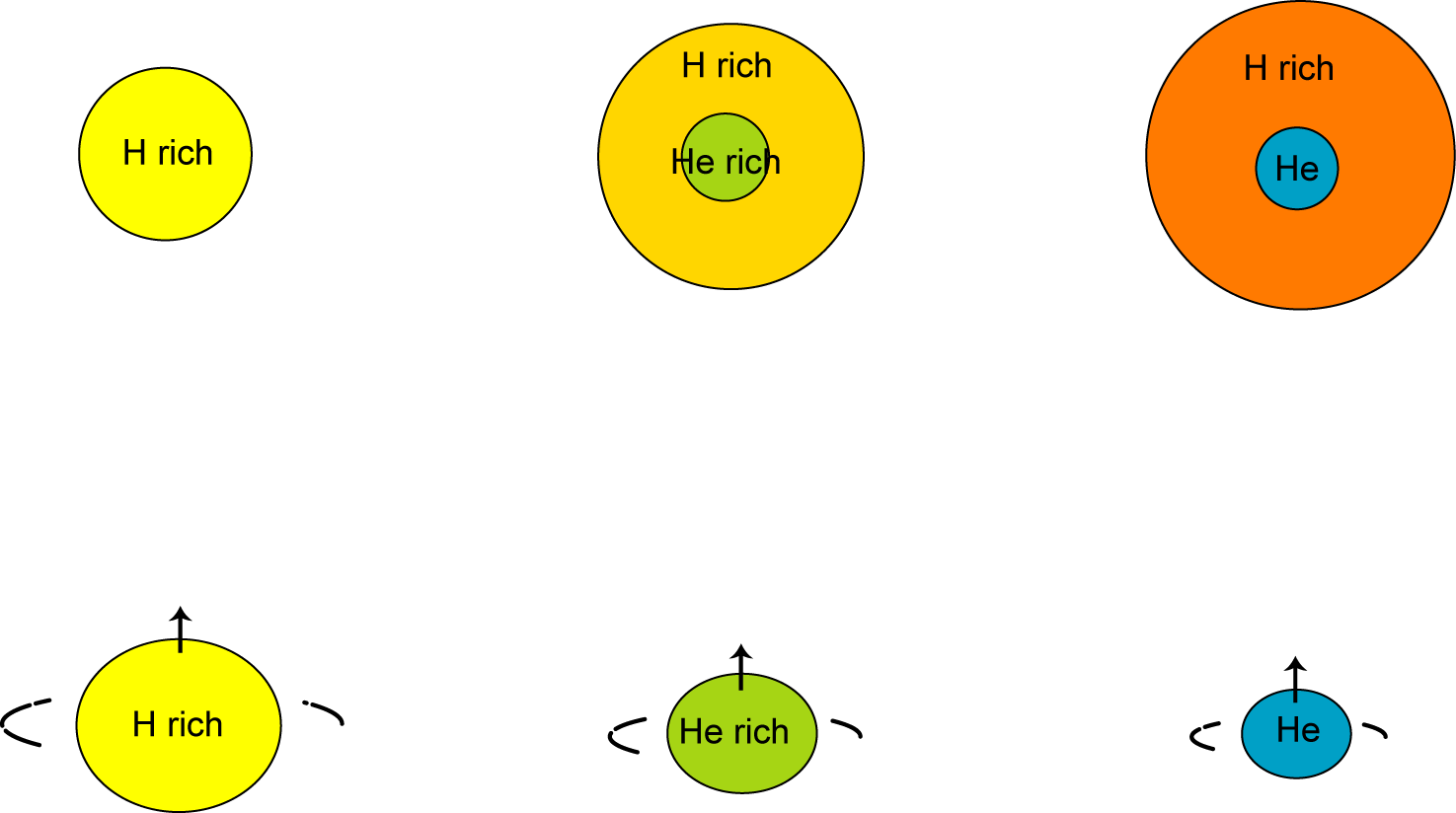}
  \caption{Cartoon representation the evolution of a core hydrogen
  burning star. A slow rotor will form a core-envelope structure (top
  row). Fast rotators will evolve quasi chemically homogeneously,
  and they will stay compact gradually becoming a WR star (bottom
  row).}\label{fig:cartoon1}
\end{figure}

\section{Surface abundances \label{sec:surf}}
The faster the initial rotation of a star of a given mass, the more
efficient is rotational mixing. For example a 20 solar mass single star
rotating with an initial equatorial rotational velocity of 180 km/s
(corresponding to a spin period of about 1.5 days) enhances its
surface nitrogen abundance by about 0.5 dex over the course of its main
sequence evolution.

Figure~\ref{fig:N} shows the surface N abundance at the end of core H
burning for the grid of single stellar models published by
\citet{Yoon+06}. In a tidally locked binary the spin period
corresponds to the orbital period. The figure therefore shows the
maximum surface abundances which we can expect for close detached
binaries. For example the sample of OB type binaries with orbital
periods ranging from 1-5 days by \citet{hilditch+05} should show
enhanced N abundances by up to 0.4~dex. Currently no spectra are
available of these systems with high enough resolution to determine
the surface abundances.  In more massive systems the enhancements will
be even larger. In principle even one well-determined binary systems
could serve as a strong test case for rotational mixing.
In fact, even if in close binaries mixing is enhanced by processes
such as tides or irradiation, these observations can be used to set an
upper limit to the efficiency of rotational mixing.

\begin{figure}[t]
\centering
  \includegraphics[width=0.7\textwidth]{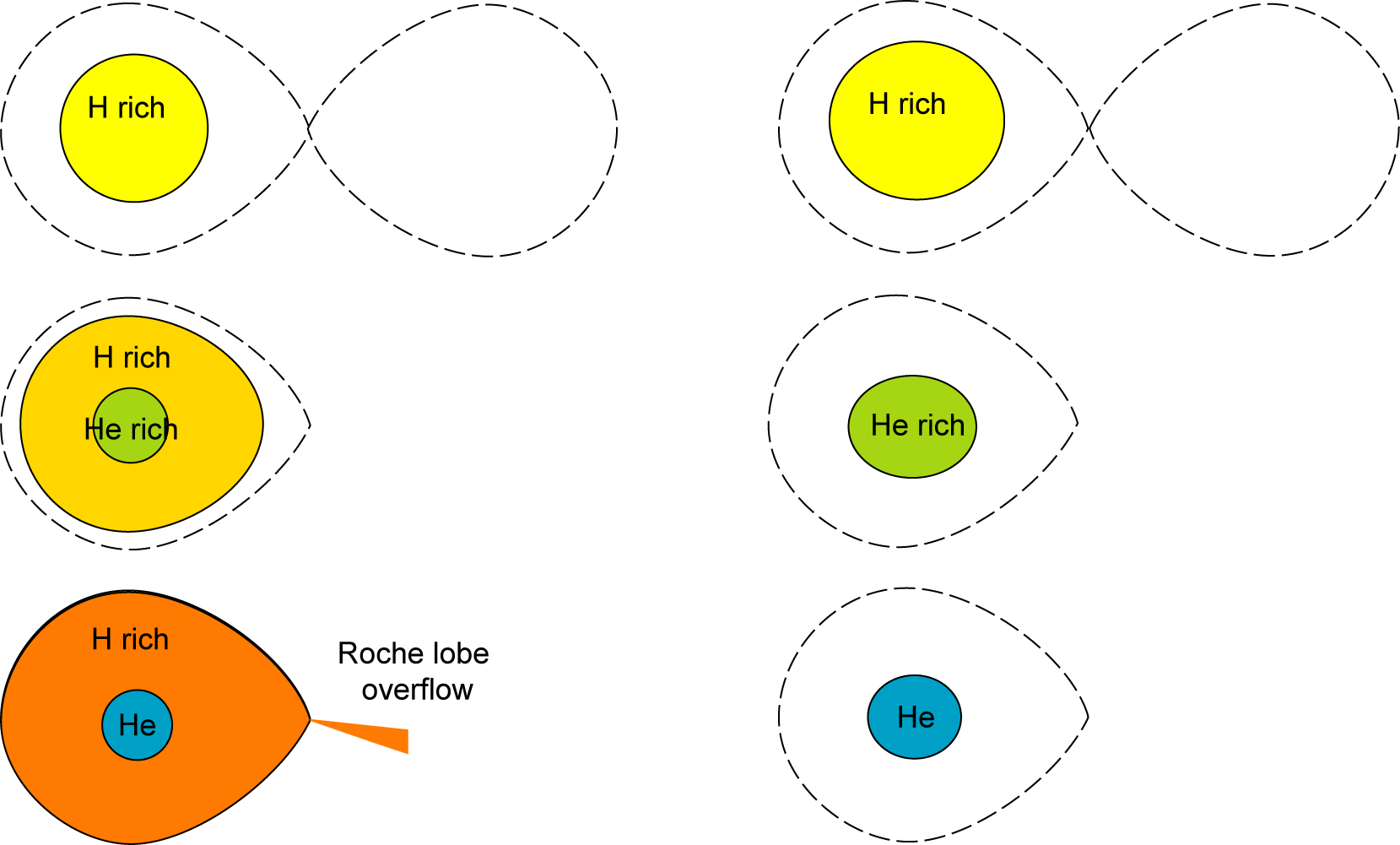}
  \caption{Cartoon of the evolution of normal star inside a Roche lobe
  (left column) and a quasi-chemically homogeneous evolving star which avoids
  Roche-lobe overflow: {\it Case M} (right column).
}\label{fig:cartoon2}
\end{figure}

\section{Chemically homogeneous evolution in binaries \label{sec:caseM}}
In rapidly rotating stars mixing can be so efficient that the stars fail
to form the usual core/envelope structure. These stars stay compact
during core H burning and gradually become WR stars \citep{Maeder87,
Yoon+Langer05}, illustrated by the cartoon in
Figure~\ref{fig:cartoon1}.  This type of evolution has been suggested
to lead to the formation of long GRB progenitors
\citep[e.g.][]{Yoon+Langer05, Cantiello+07}.

\begin{figure}[t]
  \centering
      \includegraphics[width=0.9\textwidth]{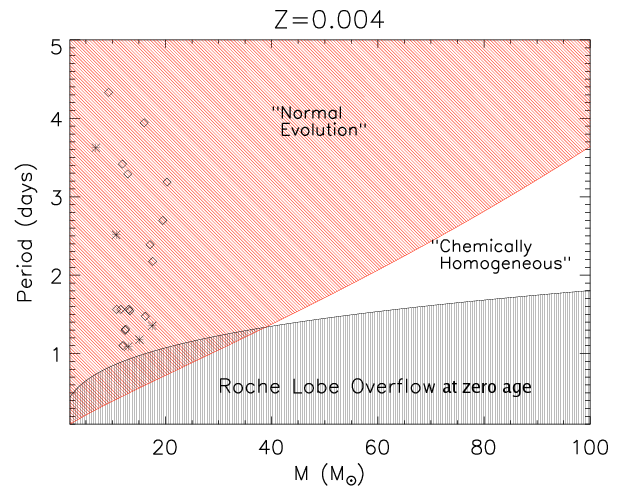}

\caption { Parameter space for binary systems in which
chemically homogeneous evolution can occur. The black hashed region is
excluded as binaries do not fit in such close orbits at zero age.  The
symbols show observed systems in the Small Magellanic Cloud
\citep{hilditch+05}.\label{fig:bin}}
\end{figure}

Figure~\ref{fig:bin} shows that for tidally locked binary systems
there is a small range in the parameter space where chemically
homogeneous evolution can occur in synchronously rotating
binaries. This implies that in such close massive binaries the stars
will stay compact and avoid Roche-lobe overflow completely (see the
cartoon in Figure~\ref{fig:cartoon2}).

\begin{figure}[t]
  \centering
  
  \includegraphics[width=\textwidth]{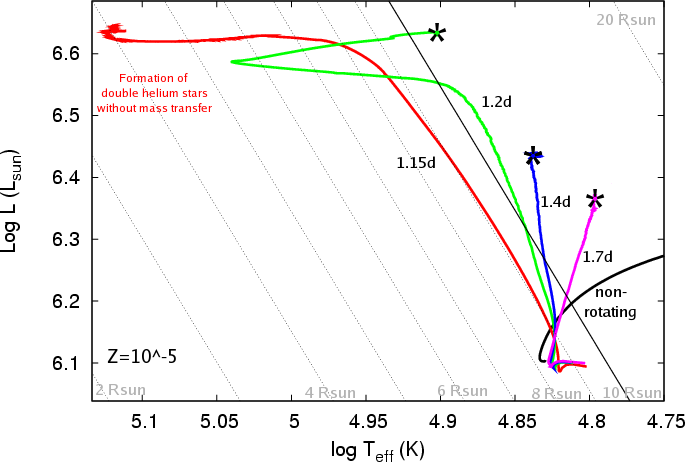}

\caption { Evolution of stars of 100~\Msun in the HR diagram, showing
the effects of rotational mixing in binaries.  The diagonal lines are
lines of constant radii.  The black line shows a non-rotating
100\Msun~single star for reference. The colored lines show the
evolution of 100\Msun~stars which orbit around a 98 \Msun mass
companion with an initial orbital period of 1.7d (purple), 1.4d
(blue), 1.2d (green), until the onset of Roche-lobe overflow (asterisk
symbol). The red curve corresponds to an initial period of 1.15d. This
system ends as two massive helium star in a close orbit.
\label{fig:hrd}}
\end{figure}

To demonstrate that this situation, predicted on the basis of single
stellar models, can actually occur in binary evolution models we
performed some preliminary calculations. Figure~\ref{fig:hrd} shows
the evolutionary track of a non-rotating 100~\Msun~star at a
metallicity of $Z = 10^{-5}$ together with four tracks corresponding
to the evolution of a 100~\Msun~star in a close binary with an equal
mass companion, in orbits with initial periods of 1.7, 1.4, 1.2
and 1.15 days, respectively.

According to non-rotating models such close systems would fill their
Roche lobe during their core hydrogen burning and start so called
Case~A mass transfer \citep{Kippenhahn+Weigert67}.
The stars in the 1.7 day and the 1.4 day systems stay more compact
than the corresponding non-rotating star, due to efficient rotational
mixing. However, they cannot avoid Roche lobe overflow before
the end of their main sequence evolution.

Although one may intuitively expect that an even closer system would
fill its Roche lobe earlier, the primary in the 1.2 day orbit system
stays compact enough to avoid mass transfer during the main
sequence. It becomes brighter at almost constant radius until central
hydrogen exhaustion. Then it contracts until the small amount of H
that is still left in the outer layers ignites. The star expands
during H shell burning and fills its Roche lobe.
In the slightly more compact system with an initial orbital period of
1.15 days, mixing is so efficient that at the end of hydrogen burning
both stars are basically pure helium stars, and Roche lobe overflow is
completely avoided.

\section {Conclusion}
Close detached binaries are potentially strong test cases to constrain
the uncertain physics of rotational mixing. Furthermore, we show that,
contrary to expectation, the very closest massive binaries could
avoid mass transfer altogether. In addition to the classic binary
cases A, B and C \citep{Kippenhahn+Weigert67, Lauterborn70}, we find a
new evolutionary scenario for very close massive binaries, which we
named Case M, where the letter M refers to the importance of mixing.  In this
case both stars are so efficiently mixed, that they remain compact and
avoid Roche-lobe overflow during the main sequence, and probably
beyond.

According to this evolutionary scenario, double helium star systems in
very close orbits can be made.  Perhaps WR20a, a binary consisting of
two detached core hydrogen burning stars of about 80 solar masses in a
3.6 day orbit \citep{Bonanos+04}, is an example of this type of
evolution.  It remains to be investigated whether this evolutionary
scenario can lead to the formation of two long GRB progenitors.


\bibliographystyle{aa}
\bibliography{references}


\end{document}